\documentclass[
showpacs,preprintnumbers,amsmath,amssymb]{revtex4}
\usepackage{amsmath}
\begin{document}

\title{CASIMIR EFFECTS NEAR THE BIG RIP SINGULARITY IN VISCOUS COSMOLOGY}

\author{Iver Brevik$^{1}$\footnote{E-mail:iver.h.brevik@ntnu.no}, Olesya Gorbunova$^{2}$\footnote{E-mail:gorbunovaog@tspu.edu.ru}, Diego S\'{a}ez-G\'{o}mez$^{3}$\footnote{E-mail: saez@ieec.uab.es}}

\medskip

\affiliation{$^{1}$Department of Energy and Process Engineering, Norwegian University
of Science and Technology, N-7491 Trondheim, Norway}

\affiliation{$^{2}$Tomsk State Pedagogical University, Tomsk, Russia}

\affiliation{$^{3}$Institut de Ciencies de l'Espai (IEEC-CSIC), E-08193 Bellaterra
(Barcelona), Spain}

 \today

\begin{abstract}

 Analytical properties of the scalar expansion in
the cosmic fluid are investigated, especially near the future
singularity, when the fluid possesses a constant bulk viscosity
$\zeta$. In addition, we assume that there is a Casimir-induced
term in the fluid's energy-momentum tensor, in such a way that the
Casimir contributions to the energy density and pressure are both
proportional to $1/a^4$, $a$ being the scale factor. A series
expansion is worked out for the scalar expansion under the
condition that the Casimir influence is small. Close to the Big
Rip singularity the Casimir term has however to fade away and we
obtain the same singular behavior for the scalar expansion, the
scale factor, and the energy density, as in the Casimir-free
viscous case.

\end{abstract}
 \maketitle
\section{Introduction}

From a hydrodynamical viewpoint it is almost surprising to notice
that the cosmic fluid - whether considered in the early or in the
late epochs - is usually taken to be nonviscous. After all, there
are two viscosity  (shear and bulk) coefficients naturally
occurring in general  linear hydrodynamics; the linear
approximation meaning physically that one is considering  only
first order deviations from thermal equilibrium. The shear
viscosity coefficient is evidently of importance when dealing with
flow near solid surfaces, but it can  be crucial also under
boundary-free conditions such as in isotropic turbulence. In later
years it has  become more common to take into account viscosity
properties of the cosmic fluid, however.  Because of assumed
spatial isotropy in the fluid the shear viscosity is usually left
out; any anisotropic deviations like those encountered in the
Kasner universe become rather quickly smoothed out. Thus only the
bulk viscosity coefficient, called $\zeta$, remains in the
energy-momentum tensor of the fluid. One should here note,
however,  that at least in the plasma region in the early universe
the value of the shear viscosity as derived from kinetic theory is
greater than the bulk viscosity by many orders of magnitude. Cf.,
for instance, Refs.~\cite{caderni77} and \cite{brevik94}.

Early treatises on viscous cosmology are given by Padmanabhan and
Chitre \cite{padmanabhan87} and Gr{\o}n \cite{gron90}, the latter
paper being an extensive review of the field up to 1990.  We have
ourselves dealt earlier with viscous entropy production in the
early universe \cite{brevik94}, and viscous fluids on the
Randall-Sundrum branes \cite{brevik04a,brevik06}. Cataldo et al.
considered viscous dark energy and phantom evolution using
Eckhart's theory of irreversible thermodynamics \cite{cataldo05}.
Kofinas et al. considered the crossing of the $w=-1$ barrier using
a brane-bulk energy exchange model with an induced gravity
curvature correction \cite{kofinas06}. As discussed by Nojiri and
Odintsov \cite{nojiri05} and by Capozziello et al.
\cite{capozziello06}, a dark fluid with a time dependent bulk
viscosity can be considered as a fluid with an inhomogeneous
equation of state. Some other recent papers on viscous cosmology
are Refs.~\cite{li09,chen09,feng09}.

A special branch of viscous cosmology is to investigate how the
bulk viscosity can influence the future singularity, commonly
called the Big Rip, when the fluid is in the phantom state
corresponding to the thermodynamic parameter $w$ being less than
$-1$.  Some recent papers  in this direction are
Refs.~\cite{brevik05,brevik05a,brevik06a,brevik08,brevik08a,brevik08c,gron08,brevik09}.
In particular, as first pointed out in Ref.~\cite{brevik05}, the
presence of a bulk viscosity proportional to the scalar expansion
$\theta$  can cause the fluid to pass from the quintessence region
into the phantom region and thereby inevitably lead to a future
singularity.

The purpose of the present paper is to generalize these viscous
cosmology theories in the sense that we take into account the {\it
Casimir effect}. We shall model the Casimir influence by writing
the total Casimir energy in the same form as for a perfectly
conducting shell, identifying the cosmological "shell" radius
essentially with the cosmic scale factor. This is a very simple,
though natural, approach to the problem. The approach is of the
same kind as that followed in an earlier quantum cosmology paper
dealing with the  expanding FRW universe in the nonviscous case
\cite{brevik00}.  We mention that there are also other ways  of
treating the influence from the Casimir effect in cosmology; the
reader may consult, for instance,
Refs.~\cite{godlowski05,schaden06,dolan92,elizalde03,elizalde06}.

As bulk viscosity corresponds to inhomogeneous Hubble rate
dependent terms in the effective equation of state, the situation
is quite similar to that of modified gravity theory. For a general
treatise on modified gravity, see the review of Nojiri and
Odintsov \cite{nojiri07}.

\section{Formalism, when the Casimir effect is omitted}

We start with the standard  FRW metric,
\begin{equation}
ds^2=-dt^2+a^2(t)d{\bf x}^2, \label{1}
\end{equation}
and set the spatial curvature $k$, as well as  the cosmological
constant $\Lambda$,  equal to zero. We let subscript 'in' refer to
 present time  quantities, and choose $t_{in}=0$. The scale factor
$a(t)$ is normalized such that $a(0) \equiv a_{in}=1$. The
equation of state is taken as
\begin{equation}
p=w\rho, \label{2}
\end{equation}
with $w$ constant. As  mentioned, $w<-1$ in the phantom region
(ordinary matter is not included in the model). The bulk viscosity
$\zeta$ is taken to be  constant. The energy-momentum tensor of
the fluid can be expressed as
\begin{equation}
 T_{\mu\nu}=\rho U_{\mu}U_{\nu}+ \tilde{p}(g_{\mu\nu}+U_{\mu}U_{\nu})\ ,
\label{3}
\end{equation}
where  $U_{\mu}$ is the comoving four-velocity  and
$\tilde{p}=p-\zeta\theta$ is the effective pressure,
$\theta=3H=3\dot{a}/a$ being the scalar expansion. The Friedmann
equations take the form
\begin{equation}
\theta^2=24\pi G\rho, \label{4}
\end{equation}
\begin{equation}
\dot{\theta}+\frac{1}{6}\theta^2=-4\pi G\tilde{p}. \label{5}
\end{equation}
The energy conservation equation leads to
\begin{equation}
\dot{\rho}+(\rho+p)\theta=\zeta \theta^2. \label{6}
\end{equation}
From the above equations the differential equation for the scalar
expansion, with $w$ and $\zeta$ as free parameters, can be derived
as
\begin{equation}
\dot{\theta}+\frac{1}{2}(1+w)\theta^2-12\pi G\zeta \theta =0.
\label{7}
\end{equation}
The solution is, when subscript zero signifies that the Casimir
effect is so far left out,
\begin{equation}
 \theta_0(t)=\frac{\theta_{in} e^{t/t_c}}{1+\frac{1}{2}(1+w)\theta_{in}t_c(e^{t/t_c}-1)}\
 .
\label{8}
\end{equation}
Here  $\theta_{in}$ is the initial (present-time) scalar expansion
and $t_c$ the 'viscosity time'
\begin{equation}
t_c=\frac{1}{12\pi G\zeta}. \label{9}
\end{equation}
Since $(1+w)<0$ by assumption, it follows from Eq.~(\ref{8}) that
the future singularity occurs at a rip time $t_{s0}$ given by
\begin{equation}
t_{s0}=t_c\ln \left[
1-\frac{2}{1+w}\frac{1}{\theta_{in}t_c}\right]. \label{10}
\end{equation}
This means that in the nonviscous limit $\zeta \rightarrow 0$,
\begin{equation}
t_{s0} =-\frac{2}{1+w}\frac{1}{\theta_{in}}, \label{11}
\end{equation}
which is independent of $\zeta$. At the other extreme, in the high
viscosity limit $t_c \rightarrow 0$,
\begin{equation}
t_{s0} = t_c\ln \left[
\frac{-2}{1+w}\frac{1}{\theta_{in}t_c}\right], \label{12}
\end{equation}
showing that $t_{s0}$ becomes small. The fluid is quickly driven
into the Big Rip singularity.

\section{The Casimir effect included}

As mentioned above, a simple and  natural way of dealing with the
Casimir effect in cosmology is to relate it to the single length
parameter in the ($k=0$) theory, namely the scale factor $a$. It
means effectively that we should put the Casimir energy $E_c$
inversely proportional to $a$. This is in accordance with the
basic property of the Casimir energy, viz. that it is a measure of
the stress in the region interior to the "shell" as compared with
the unstressed region on the outside. The effect is evidently
largest in the beginning of the universe's evolution, when $a$ is
small. At late times, when $a\rightarrow \infty$, the Casimir
influence should be expected to fade away. As we have chosen $a$
nondimensional, we shall introduce an auxiliary length $L$ in the
formalism. Thus we adopt in model in which
\begin{equation} E_c=\frac{C}{2La}, \label{13}
\end{equation}
where $C$ is a nondimensional constant. This is the same form as
 encountered for the case of a perfectly conducting shell
 \cite{milton78}. In the last-mentioned case, $C$ was found to
 have the value
\begin{equation}
C=0.09235. \label{14}
\end{equation}
The expression (\ref{13}) is  of the same form as adopted in
Ref.~\cite{brevik00} (cf. also \cite{brevik01}). It is strongly
related to the assumptions made by Verlinde when dealing with the
holographic principle in the universe \cite{verlinde00}. Cf. also
the papers \cite{brevik02} and \cite{brevik02a} dealing with the
holographic principle applied to viscous cosmology.

In the following we shall for definiteness assume $C$ to be
positive, corresponding to a repulsive Casimir force, though $C$
will not necessarily be required to have the value (\ref{14}). The
repulsiveness is a characteristic feature of conducting shell
Casimir theory, following from electromagnetic field theory under
the assumption that dispersive short-range effects are left out
(\cite{milton78}; cf. also \cite{brevik01}). Another assumption
that we shall make, is that $C$ is small compared with unity. This
is physically reasonable, in view of the conventional feebleness
of the Casimir force.

The expression (\ref{13}) corresponds to a Casimir pressure
\begin{equation}
p_c=\frac{-1}{4\pi (La)^2}\frac{\partial E_c}{\partial
(La)}=\frac{C}{4\pi L^4a^4}, \label{15}
\end{equation}
and leads consequently to a Casimir energy density $\rho_c \propto
1/a^4$.

The Casimir energy-momentum tensor
\begin{equation}
T_{\mu\nu}^c=\rho_c U_\mu U_\nu+p_c(g_{\mu\nu}+U_\mu U_\nu),
\label{16}
\end{equation}
together with the Casimir equation of state $p_c=w_c\rho_c$, yield
the energy balance
\begin{equation}
\frac{\dot{\rho}_c}{\rho_c}+(1+w_c)\theta=0, \label{17}
\end{equation}
having the solution $\rho_ca^{3(1+w_c)}=$ constant. To get $\rho_c
\propto 1/a^4$ we must have $w_c=1/3$. The Casimir contributions
to the pressure and energy density become accordingly
\begin{equation}
p_c=\frac{C}{8\pi L^4a^4}, \quad \rho_c=\frac{3C}{8\pi L^4a^4}.
\label{18}
\end{equation}
The Friedmann equations now become
\begin{equation}
\theta^2=24\pi G\left( \rho+ \frac{3C}{8\pi L^4a^4}\right),
\label{19}
\end{equation}
\begin{equation}
\dot{\theta}+\frac{1}{6}\theta^2=-4\pi G\left( w\rho -\zeta \theta
+\frac{C}{8\pi L^4a^4}\right), \label{20}
\end{equation}
while the energy conservation equation preserves its form,
\begin{equation}
\dot{\rho}+(1+w)\rho \theta =\zeta \theta^2. \label{21}
\end{equation}
Note again that  we are considering the {\it dark energy fluid}
only, with density $\rho$ and thermodynamical parameter $w$. The
ordinary matter fluid (dust) is left out.

Solving $\rho$ from Eq.~(\ref{19}) and inserting into
Eq.~(\ref{21}) we obtain as governing equation for the scalar
expansion
\begin{equation}
\dot{\theta}+\frac{1}{2}(1+w)\theta^2-12\pi G\zeta \theta = -
(1-3w)\frac{3GC}{2L^4a^4}\ . \label{22}
\end{equation}
It is convenient to introduce the constant $\alpha$, defined as
\begin{equation}
\alpha=-(1+w)>0, \label{23}
\end{equation}
and also to define the quantity $X(t)$,
\begin{equation}
X(t)=1-\frac{1}{2}\alpha \theta_{in}t_c(e^{t/t_c}-1), \label{24}
\end{equation}
which satisfies
\begin{equation}
X(0)=1, \quad X(t_{s0})=0. \label{25}
\end{equation}
In view of the assumed smallness of $C$ we now make a Stokes
expansion for $\theta$ to the first order,
\begin{equation}
\theta(t)=\theta_0(t)+C\theta_1(t)+O(C^2), \label{26}
\end{equation}
using henceforth the convention that subscript zero refers to the
$C=0$ case. The zeroth order solution is
\begin{equation}
\theta_0(t)=\theta_{in}X^{-1}e^{t/t_c}, \label{27}
\end{equation}
in accordance with Eq.~(\ref{8}). It corresponds to the zeroth
order scale factor
\begin{equation}
a_0(t)=X^{-\frac{2}{3\alpha}}, \label{28}
\end{equation}
satisfying $a_0(0)=1$ as before.

As the right hand side of Eq.~(\ref{22}) is already of order $C$,
we can replace $a(t)$ with $a_0(t)$ in the denominator. Thus we
get the following equation for the first order correction
coefficient $\theta_1$:
\begin{equation}
\dot{\theta}_1-\left(\alpha \theta_{in}X^{-1}e^{t/t_c}+12\pi
G\zeta
\right)\theta_1=-(1-3w)\frac{3G}{2L^4}X^{\frac{8}{3\alpha}}.
\label{29}
\end{equation}
We impose the same initial condition for the scalar expansion as
in the $C=0$ case: $\theta(0)=\theta_0(0) \equiv \theta_{in}$. It
means according to Eq.~(\ref{26}) that $\theta_1(0)=0$.

The homogeneous solution of Eq.~(\ref{29}), called $\theta_{1h}$,
may be written
\begin{equation}
\theta_{1h}(t)=\exp \left[ \int_0^t(\alpha
\theta_{in}X^{-1}e^{t/t_c}+12\pi G\zeta)dt \right], \label{30}
\end{equation}
satisfying $\theta_{1h}(0)=1$. The the full solution becomes
\begin{equation}
\theta_1(t)=-(1-3w)\frac{3G}{2L^4}\theta_{1h}(t)\cdot \int_0^t
\frac{X^{\frac{8}{3\alpha}}}{\theta_{1h}} dt, \label{31}
\end{equation}
satisfying $\theta_1(0)=0$. The two terms on the right hand side
of Eq.~(\ref{26}), $\theta_0(t)$ and $C\theta_1(t)$, are
accordingly determined.

\subsection{ Analytic approximation for low viscosity}

 Although in general the expression for $\theta(t)$ has to be calculated
numerically, the main features of the solution can be shown
already analytically. Consistent with the assumed smallness of $C$
we need not distinguish between the rip time $t_s$ corresponding
to $C\neq 0$ and the rip time $t_{s0}$ corresponding to $C=0$. Let
us assume for mathematical simplicity the low-viscosity limit
\begin{equation}
\theta_{in}t_c \gg 1, \label{32}
\end{equation}
being physically  the most important case also. It corresponds to
$t_{s0}/t_c =2/(\alpha \theta_{in}t_c) \ll 1$. Then,
\begin{equation}
X(t)\sim \frac{1}{2}\alpha \theta_{in}(t_{s0}-t)=\frac
{t_{s0}-t}{t_{s0}}, \label{33}
\end{equation}
\begin{equation}
\theta_0(t) \sim \frac{2}{\alpha}\,\frac{1}{t_{s0}-t}, \label{34}
\end{equation}
\begin{equation}
a_0(t) \sim \left(
\frac{t_{s0}}{t_{s0}-t}\right)^{\frac{2}{3\alpha}}. \label{35}
\end{equation}
Both $\theta_0(t)$ and $a_0(t)$ diverge ($\alpha >0$ by
assumption). Using Eq.~(\ref{33})
 we can calculate $\theta_1(t)$ from Eq.~(\ref{31}),
\begin{equation}
\theta_1(t) =-(1-3w)\frac{9G}{2L^4}\,\frac{\alpha
t_{s0}}{8+9\alpha}\,\frac{1-(1-t/t_{s0})^{\frac{8}{3\alpha}+3}}{(1-t/t_{s0})^2}.
\label{36}
\end{equation}
From the expansion (\ref{26}) we thus obtain for the scalar
expansion to the first order in $C$,
\begin{equation}
\theta(t)=\frac{2}{\alpha
t_{s0}}\frac{1}{1-t/t_{s0}}\left\{1-(1-3w)\frac{9GC}{4L^4}\,\frac{\alpha^2t_{s0}^2
}{8+9\alpha}\,\frac{1-(1-t/t_{s0})^{\frac{8}{3\alpha}+3}}{1-t/t_{s0}}\right\}.
\label{37}
\end{equation}
The viscosity is absent in this expression. This is as we would
expect in view of the low-viscosity approximation.

The expression (\ref{37}) cannot, however, be valid near the
singularity. The reason is that the Taylor expansion in $C$ in
Eq.~(\ref{26}) is not applicable at $t=t_{s0}$. The solution
(\ref{37}) can  be applied safely as long as $t$ stays
considerably smaller than $t_{s0}$.  By making a first order
expansion in $t/t_{s0}$ of the expression between the curly
parentheses  we can write the solution in simplified form as
\begin{equation}
\theta(t)=\frac{2}{\alpha
t_{s0}}\frac{1}{1-t/t_{s0}}\left\{1-(1-3w)\frac{3GC\alpha
t_{s0}}{4L^4}t\right\}, \quad t \ll t_{s0}. \label{38}
\end{equation}
As $(1-3w)>0$ this means that $\theta(t)$ becomes slightly reduced
by the Casimir term. The repulsive Casimir force causes the energy
density $\rho$ in Eq.~(\ref{19}) to be slightly smaller than in
the $C=0$ case.

To deal with the conditions close to the singularity, we have to
go back to the governing equations themselves.

\subsection{Behavior close to the singularity}

To begin with, it is instructive to  list the general
classification of possible future singularities as given in
Refs.~\cite{nojiri05a} and \cite{brevik09}. If $t_s$ denotes the
rip time, one has four types,

(i) Type I (original "Big Rip"): For $t \rightarrow t_s,~
a\rightarrow \infty,~\rho \rightarrow \infty$, and $|p|\rightarrow
\infty$, or $p$ and $\rho$ are finite at  $t=t_s$.

(ii) Type II ("sudden"): For $t\rightarrow t_s,~a\rightarrow
a_s,~\rho \rightarrow \rho_s$, and $|p|\rightarrow \infty$,

(iii) Type III: For $t\rightarrow t_s,~a\rightarrow a_s,~\rho
\rightarrow \infty$, and $|p|\rightarrow \infty$,

(iv) Type IV: For $t\rightarrow t_s,~a\rightarrow a_s,~\rho
\rightarrow 0,~|p|\rightarrow 0$, or $p$ and $\rho$ are finite.
Higher order derivatives of $H$ diverge.

\bigskip

The singularities we have been contemplating above are seen to be
of Type I. As $a\rightarrow \infty$ near the singularity, we can
draw the important conclusion from Friedmann's equations
(\ref{19}) and (\ref{20}) that the influence from the Casimir term
fades away. Close to the singularity we simply obtain the same
solutions as in the Casimir-free case.

Let now $t_{\zeta s}$ denote the singularity time in the presence
of viscosity. We thus get
\begin{equation}
t_{s\zeta}=t_c\ln
\left(1+\frac{2}{\alpha}\frac{1}{\theta_{in}t_c}\right).
\label{39}
\end{equation}
It corresponds to $\theta(t_{s\zeta})=\infty$. We see that
$t_{s\zeta}$ is always less than the singularity time for the
nonviscous case,
\begin{equation}
t_{s\zeta}<t_{s,\zeta=0}=\frac{2}{\alpha \,\theta_{in}}.
\label{40}
\end{equation}
For the scalar expansion we find close to the singularity, again
assuming for simplicity low viscosity so that $\theta_{in}t_c \gg
1$ \cite{brevik08a},
\begin{equation}
\theta(t)=\frac{2/\alpha}{t_{s\zeta}-t}, \quad t\rightarrow
t_{s\zeta}. \label{41}
\end{equation}
In turn, this corresponds to
\begin{equation}
a(t) \sim \frac{1}{(t_{s\zeta}-t)^{2/3\alpha}}, \quad t\rightarrow
t_{s\zeta}, \label{42}
\end{equation}
\begin{equation}
\rho(t)\sim \frac{1}{(t_{s\zeta}-t)^2}, \quad t\rightarrow
t_{s\zeta}. \label{43}
\end{equation}

\subsection{On the nonviscous case}

It may finally be worthwhile to consider the entirely nonviscous
case, while keeping $C>0$. Setting $\zeta=0$ we get from
Eq.~(\ref{22}) the governing  equation for the scale factor $a$:
\begin{equation}
a^3\ddot{a}+\frac{1}{2}(1+3w)a^2{\dot{a}}^2=-(1-3w)\frac{GC}{2L^4}.
\label{44}
\end{equation}
This equation is still not solvable analytically. We get for
$X(t), \theta_0(t)$ and $a_0(t)$ the same expressions as in
Eqs.~(\ref{33}) - (\ref{35}). Similarly, we get for $\theta(t)$
the same expansion (\ref{37}) as before, with $t_{s0}=2/(\alpha
\theta_{in})$. These results are as expected, in view of
 the property of continuity with respect to variation of parameters.

\section{Concluding remarks}

Considerable attention has recently been devoted to the behavior
of the dark energy fluid near the future singularity. Various
possibilities have been contemplated. In addition to the
references given above \cite{godlowski05} -\cite{elizalde06}, we
may refer also to the papers \cite{caldwell03}. It has even been
suggested that the future singularity can be avoided via quantum
gravity effects. Thus in Ref.~\cite{elizalde04} it is shown how
the universe may turn into a de Sitter phase.

\vspace{0.3cm}

Let us finally summarize our results above:

\vspace{0.3cm}

1)  If $\zeta>0$ and $C=0$, the rip singularity time $t_{s0}$ is
given by Eq.~(\ref{10}). In particular, if $\zeta \rightarrow 0$
then Eq.~(\ref{11}) holds.

2) If $\zeta>0$ and $C>0$, the scalar expansion $\theta(t)$ is
given by the first-order series (\ref{26}), with $\theta_0(t)$ and
$\theta_1(t)$ given by Eqs.~(\ref{27}) and (\ref{31}). In the
low-viscosity case $\theta_{in}t_c \gg 1$, $\theta(t)$ is given by
the series (\ref{37}) which, however, is not applicable near the
singularity as $\theta(t)$ is not analytic in $C$ at the
singularity.

3) Near the singularity, the Casimir effect fades away and the
viscous rip time $t_{s\zeta}$ is given by Eq.~(\ref{39}).
Corresponding values for $\theta(t),a(t)$ and $\rho(t)$ near the
singularity follow from Eqs.~(\ref{41}) - (\ref{43}).

4)  If $\zeta=0$ and $C>0$, the governing equation for $a(t)$ is
Eq.~(\ref{44}).

\bigskip

{\bf Acknowledgment}

\bigskip

 O. Gorbunova acknowledges support from the ESF Programme "New
Trends and Applications of the Casimir Effect" and by Grant the
Scientific School LRSS Project N.2553.2008.2. D. Saez-Gomez
acknowledges a grant from  MICINN (Spain), project FIS2006-02842,
DSG's research was performed while on leave at Department of
Physics of NTNU (Norway)


\begin{thebibliography}{99}
\bibitem{caderni77}
N. Caderni and R. Fabbri, Phys. Lett. {\bf 69B}, 508 (1977).
\bibitem{brevik94}
I. Brevik and L. T. Heen, Astrophys. Space Sci. {\bf 219}, 99
(1994).
\bibitem{padmanabhan87}
T. Padmanabhan and S. M. Chitre, Phys. Lett. A {\bf 120}, 433
(1987).
\bibitem{gron90}
{\O}. Gr{\o}n, Astrophys. Space Sci. {\bf 173}, 191 (1990).
\bibitem{brevik04a}
 I. Brevik and A. Hallanger, Phys. Rev. D {\bf 69}, 024009 (2004).
 \bibitem{brevik06}
 I. Brevik, J. M. B{\o}rven and S. Ng, Gen. Relativ. Gravit. {\bf
 38}, 907 (2006).
 \bibitem{cataldo05}
 M. Cataldo, N. Cruz and S. Lepe, Phys. Lett. B {\bf 619}, 5
 (2005).
 \bibitem{kofinas06}
 G. Kofinas, G. Panotopoulos and N. Tomaras, JHEP {\bf 0601}, 107
 (2006).
 \bibitem{nojiri05}
 S. Nojiri and S. D. Odintsov, Phys. Rev. D {\bf 72}, 023003
 (2005).
 \bibitem{capozziello06}
 S. Capozziello, V. F. Cardone, E. Elizalde, S. Nojiri and S. D.
 Odintsov, Phys. Rev. D {\bf 73}, 043512 (2006).
 \bibitem{li09}
 B. Li and J. D. Barrow, arXiv:0902.3163 [gr-qc].
 \bibitem{chen09}
 J. Chen and Y. Wang, arXiv:0904.2808 [gr-qc].
 \bibitem{feng09}
 C. J. Feng and Xin-Zhou Li, arXiv:0905.0527 [astro-ph.CO].
 \bibitem{brevik05}
 I. Brevik and O. Gorbunova, Gen. Relativ. Gravit. {\bf 37},
 2039 (2005).
 \bibitem{brevik05a}
 I. Brevik, O. Gorbunova  and Y. A. Shaido, Int. J. Mod. Phys. D
 {\bf 14}, 1899 (2005).
 \bibitem{brevik06a}
 I. Brevik, Gen. Rel. Gravit. {\bf 38}, 1317 (2006).
 \bibitem{brevik08}
 I. Brevik, Gravitation and Cosmology {\bf 14}, 332 (2008).
 \bibitem{brevik08a}
 I. Brevik and O. Gorbunova, Eur. Phys. J. C {\bf 56}, 425 (2008).
 \bibitem{brevik08c}
 I. Brevik, Eur. Phys. J. C {\bf 56}, 579 (2008).
 \bibitem{gron08}
 {\O}. Gr{\o}n, in {\it The Casimir Effect and Cosmology: A volume
 in honour of Professor Iver H. Brevik on the occasion of his 70th
 birthday}, Editors S. D. Odintsov, E. Elizalde and O. G.
 Gorbunova (Tomsk State Pedagogical University Press, 2008), p.
 75; arXiv:0812.2549.
 \bibitem{brevik09}
 I. Brevik and O. Gorbunova, in {\it The Problems of Modern
 Cosmology: A volume in honour of Professor S. D. Odintsov on the
 occasion of his 50th birthday}, Editor P. M. Lavrov (Tomsk State
 Pedagogical University Press, 2009), p. 106; arXiv:0811.1129.
 \bibitem{brevik00}
 I. Brevik, K. A. Milton, S. D. Odintsov and K. E. Osetrin, Phys.
 Rev. D {\bf 62}, 064005 (2000).
 \bibitem{godlowski05}
 W. Godlowski, M. Szydlowski and Z. H. Zhu, Gravitation and
 Cosmology {\bf 11}, 1 (2005).
\bibitem{schaden06}
M. Schaden, Phys. Rev. A {\bf 73}, 042102 (2006).
\bibitem{dolan92}
B. P. Dolan and C. Nash, Commun. Math. Phys. {\bf 148}, 139
(1992).
\bibitem{elizalde03}
E. Elizalde, S. Nojiri, S. D. Odintsov and S. Ogushi, Phys. Rev. D
{\bf 67}, 063515 (2003).
\bibitem{elizalde06}
E. Elizalde, J. Phys. A {\bf 39}, 6299 (2006).
\bibitem{nojiri07}
S. Nojiri and S. D. Odintsov, Int. J. Geom. Meth. Mod. Phys. {\bf
4}, 115 (2007) [arXiv:hep-th/0601213].
\bibitem{milton78}
 K. A. Milton, L. L. DeRaad, Jr. and J. Schwinger, Ann. Phys.
 (N.Y.) {\bf 115}, 388 (1978).
 \bibitem{brevik01}
 I. Brevik, B. Jensen and K. A. Milton, Phys. Rev. D {\bf 64},
 088701 (2001).
 \bibitem{verlinde00}
 E. Verlinde, arXiv:hep-th/0008140.
 \bibitem{brevik02}
 I. Brevik and S. D. Odintsov, Phys. Rev. D {\bf 65}, 067302
 (2002).
 \bibitem{brevik02a}
 I. Brevik, Phys. Rev. D {\bf 65}, 127302 (2002).
\bibitem{nojiri05a}
 S. Nojiri, S. D. Odintsov and S. Tsujikawa, Phys. Rev. D {\bf
 71}, 063004 (2005).
 \bibitem{caldwell03}
 R. R. Caldwell, M. Kamionkowski and N. N. Weinberg, Phys. Rev. Lett.
 {\bf 91}, 071301 (2003); S. Nojiri and S. D. Odintsov, Phys.
 Lett. B {\bf 562}, 147 (2003).
\bibitem{elizalde04}
 E. Elizalde, S. Nojiri and S. D. Odintsov, Phys. Rev. D {\bf 70}, 043539
 (2004).





\end{thebibliography}
\end{document}